\documentclass[aps,preprint]{revtex4}
\usepackage[dvips]{graphicx}
\begin{document}

\title
{
Switching magnetization of nano-scale ferromagnetic particle   
using non-local spin injection 
}

\author
{ 
T. Kimura, Y. Otani
}
\address{
Institute for Solid State Physics, University of Tokyo, 
5-1-5 Kashiwanoha, Kashiwa, Chiba 277-8581, Japan \\
RIKEN FRS, 2-1 Hirosawa, Wako, Saitama 351-0198, Japan \\
CREST, JST, Honcho 4-1-8, Kawaguchi, Saitama, 332-0012, Japan 
}

\author
{ 
J. Hamrle
}
\address{
RIKEN FRS, 2-1 Hirosawa, Wako, Saitama 351-0198, Japan \\
CREST, JST, Honcho 4-1-8, Kawaguchi, Saitama, 332-0012, Japan 
}


\date{\today}
\begin{abstract}
We have performed non-local spin injection into a nano-scale ferromagnetic particle 
configured in a lateral spin valve structure 
to switch its magnetization only by spin current.  
The non-local spin injection aligns the magnetization of the particle 
parallel to the magnetization of the spin injector.   
The responsible spin current for switching is estimated from the experiment 
to be about 200 $\mu$A , which is reasonable 
compared with the values  obtained for conventional pillar structures.  
Interestingly the switching always occurs from anti-parallel to parallel in the particle/injector 
magnetic configurations, 
whereas no opposite switching is observed.  
Possible reasons for this discrepancy are discussed.  
\end{abstract}

\maketitle
Spin-dependent transport properties have drawn enormous attention 
owing to the novel idea to utilize the spin angular momentum to operate future spintronic devices 
including a magnetic random access memory.\cite{Bauer}  
Unlike conventional inductive recording methods, the spin angular momentum ({\it spin torque}) of 
conduction electrons is now employed  to switch the magnetization. 
The current-induced magnetization reversal 
becomes one of the key technologies for developing spintronic devices.  
The switching mechanism due to spin torque is explained with 
a model proposed by Slonczewski in 
which the torque exerted on the magnetization is proportional to 
the injected spin current.  
This clearly indicates that the spin current is essential to realize the magnetization switching due to the spin injection.  
Most of the present spin-transfer devices consist of vertical multilayered nanopillars 
in which typically two magnetic layers are separated by a nonmagnetic metal layer.\cite{CPP2,CPP3,CPP4}  
In such vertical structures, the charge current always flows together with the spin current, 
thereby undesirable Joule heat is generated.  
Our recent experiments have demonstrated that the spin currents are effectively absorbed 
into an additionally connected metallic wire with a small spin resistance.\cite{Kimura1, Kimura2}
This implies that the spin current without a charge flow can be selectively injected 
into a ferromagnetic particle with a small spin resistance such as a Permalloy particle, 
replaced with the wire, and may contribute to the spin torque.  
To test this idea, a nano-scale ferromagnetic particle is configured for 
a lateral non-local spin injection device as in Figs. 1(a) and (b).

Lateral multi-terminal spin-injection device used for this study is fabricated 
by means of the conventional lift-off techniques.  
Figures 1(c) and 1(d) show the scanning-electron-microscope (SEM) images of a fabricated device.  
The device consists of a large Permalloy (Py) pad 30 nm in thickness, 
a Cu cross 100 nm in width and 80 nm in thickness, and a Py nano-scale particle, 
50 nm in width, 180 nm in length and 6 nm in thickness.  
A gold wire 100 nm in width and 40 nm in thickness is connected 
to the Py particle to reduce the effective spin resistance, 
resulting in high spin current absorption into the Py particle.\cite{Kimura2}  
The magnetic field is applied along the easy axis of the Py particle.  
We note here that the dimensions of Py pad and Cu wires are  chosen large so that 
the charge current up to 15 mA can flow through them.  
Py layer is grown using an electron beam evaporator 
with a base pressure of 2 $\times 10^{-9}$ Torr.   
The Cu and Au wires are evaporated by a resistance heating evaporator 
with a base pressure of 3 $\times 10^{-8}$ Torr.  
The interface between the Py and Cu and  that between the Py and Au 
are well cleaned by Ar-ion milling prior to each deposition.   
Very low resistance of the interface assures  good ohmic contact.  
The distance between the Py pad and the particle is 400 nm.  
The resistivities of Py and Cu wires are respectively 10.2 $\mu\Omega$cm 
and 1.14 $\mu\Omega$cm at 77 K.  
All the measurements are performed at 77 K by using conventional lock-in technique.

Before showing the experimental results, we discuss spin-current absorption 
into the electrically floating additional wire.  
We have demonstrated that the spin current distribution can be calculated by 
the model base on the spin-resistance circuit 
and that the spin current favors to flow in the subsection which has small spin resistance.\cite{Kimura1, Kimura2}  
The spin resistance is given by $2 \rho_i \lambda_i /(1-\alpha_i^2)S_i$, 
where $\rho_i$, $\lambda_i$, $\alpha_i$ and $S_i$ are 
the resistivity, spin diffusion length, spin polarization and the cross section 
of the layer.  
For example, the spin resistance $R_{S}^{\rm Cu}$ for the Cu wire 
with the cross section of 100 nm $\times$ 80 nm can be calculated as 2.85 $\Omega$.  
Here, we took the value of 1 $\mu$m for the spin diffusion length of Cu wire 
obtained in our experiments.\cite{Kimura3}  
The effective cross section for the spin resistance of the Py particle 
in Fig.\ 1(d) is given by the junction area between the Py particle and the Cu wire 
because the spin diffusion length of the Py is quite short.  
Therefore, we obtain the spin resistance $R_{S}^{\rm Py}$ for the Py particle as 0.08 $\Omega$.  
Here, we use the spin polarization $\alpha_{\rm Py} = 0.2$ and spin diffusion length 
$\lambda_{\rm Py} = 2$ nm obtained in our experiments.\cite{Kimura3}   
Important is that $R_{S}^{\rm Py}$ is about 2 orders of magnitude smaller than $R_{S}^{\rm Cu}$.  
Thus, the spin current favors to be absorbed into the Py particle 
although the particle is electrically floating.  
Such an absorption can be employed as a method to inject the spin current non-locally.

To begin with, the non-local spin valve (NLSV) measurements 
are performed to understand the spin accumulation behaviors in our devices.  
Figure 2(a) shows the NLSV signal with the inset of the measurement probe configuration.  
We can see a clear spin signal of 0.18 m$\Omega$.  
Here, the resistance changes at low and high fields 
correspond to the relative magnetic switching of the Py pad and particle, 
from parallel (P) to anti-parallel (AP) states and {\it vice versa}.  
These results prove that the spin current from the Py pad is absorbed into the Py particle.  
It should be noted that the measured spin signal is smaller than the value 
expected from our previous experiments.\cite{Kimura1, Kimura2, Kimura3}  
As will be mentioned later, this is because the Py particle used in the present study has 
smaller spin resistance than the values estimated from our lateral spin-valve experiments.

Then, we examine the effect of the non-local spin injection into the Py particle 
with using the same probe configuration.  
Before performing the non-local spin injection, the magnetization configuration is set 
in the AP configuration by controlling the external magnetic field.  
The non-local spin injection is performed by applying large pulsed currents up to 15 mA 
with the same current probes for the NLSV measurement in the absence of magnetic field.  
Note that the current pulse is triangular shape with the duration of 1s.  
After the non-local spin injection, the NLSV signal is 
successively measured to determine the magnetic state of the Py particle.  
In this way, as shown in Fig.\ 2(b), the NLSV signal after the non-local spin injection 
as a function of the amplitude of the pulsed current is obtained.  
When the magnitude of the pulsed current is increased positively in the AP state, 
no signal change is observed up to 15 mA.  
On the other hand, for the negative scan, the abrupt signal change 
is observed at $-14$ mA.  
The change in resistance at $-14$ mA is 0.18 m$\Omega$, 
corresponding to that of the transition from AP to P states.  
The magnetization direction of the Py particle 
is confirmed to be parallel to the Py pad by sweeping the magnetic field 
with measuring the NLSV signal.  
After the transition from the AP to P states, 
the current is positively increased in the P state.  
However, we observe no signal change 
even though the amplitude of the pulsed current was increased up to 15 mA.  
In this measurement, there are 2 equivalent AP states in which 
the magnetization of the Py particle directs towards either 
left or right in Fig. 2 (b).    
Both AP states are found to transform to the P state in the same manner.  
We can exclude as follows a possibility that the magnetization of the Py particle is switched 
by the current-induced Oersted field.  
In the probe configuration for the non-local spin injection, 
the charge current passes through the Cu cross and induces the Oersted field.  
However, the induced field is normal to the substrate 
and thus does not switch the magnetization of the Py particle 
since the demagnetizing field of nearly 1 T is far bigger than the Oersted field.

Figure 3(a) shows the similar measurement performed with the different probe configuration 
in the inset.  
We also obtain a clear spin signal of 0.19 m$\Omega$, 
slightly larger than that in the previous configuration.  
The discrepancy between the two 
is considered due to the inhomogeneous spin current distribution.  
Then, the similar spin injection measurements are performed.   
Figure 3(b) shows the NLSV signal after the non-local spin injection 
as a function of the amplitude of the pulsed current.  
The clear transition from the AP to P states is observed 
whereas the reverse P to AP transition is not observed.  
In this case, the switching occurs at $-13.3$ mA slightly smaller than 
that in the previous probe configuration.  
This is because the larger spin accumulation at the interface induces the larger spin current 
than in the previous configuration.  
In this probe configuration, the distribution of the Oersted field 
is different from the previous experiment.  
No remarkable difference in the transition behaviors between Figs. 2 and 3 
supports that the observed AP to P transition is not originated by the Oersted field.  
We like to point here that for both cases the Oersted field exerted normal to the substrate 
causes a deviation from the collinear magnetic configuration between Py pad and particle\cite{Kent},   
and may assist the magnetization switching of the Py particle due to the spin torque, 
leading to the reduction of the switching current.

We estimate the magnitude of injected spin current into the Py particle in the AP state 
When the electrons are injected from the Py pad into the Cu wire (negative current), 
the Cu wire is magnetized in parallel to the Py pad due to the spin accumulation.  
The continuity of the chemical potential at the interface also brings about 
the spin splitting in the Py particle, as shown in Figs.\ 4(a) and 4(c).  
The spin-dependent chemical potential of the Py particle in anti-prallel to 
the Py pad is given by\cite{Takahashi, vanSon} 
\begin{eqnarray}
\mu_{\uparrow} & = &  \mu_{\rm i} \left ( \frac{\alpha_{\rm Py}}{2} - 
\frac{1 +  \alpha_{\rm Py}}{2} e^{-\frac{x}{\lambda_{\rm Py}}} \right) .\\ 
\mu_{\downarrow} & = &  \mu_{\rm i} \left ( \frac{\alpha_{\rm Py}}{2}  + \frac{1 
- \alpha_{\rm Py}}{2} e^{-\frac{x}{\lambda_{\rm Py}}} \right) .
\end{eqnarray}
Defining the spin current as $I_S = I_{\uparrow} - I_{\downarrow} = 
-(S_{\rm Py} \sigma_{\uparrow}/e) (\partial \mu_{\uparrow}/\partial x) + 
(S_{\rm Py} \sigma_{\downarrow}/e) (\partial \mu_{\downarrow}/\partial x) $ 
yields the injected (absorbed) total spin current into the Py particle : 
\begin{eqnarray}
I_{S}   & = & \frac{\sigma_{\rm Py} S}{e} \left( \frac{(1 - \alpha_{\rm Py})}{2} 
\left. \frac{\partial \mu_\uparrow}{\partial x} \right|_{x=0} - 
 \frac{(1 + \alpha_{\rm Py})}{2}  \left. \frac{\partial \mu_\downarrow}{\partial x} \right|_{x=0}  \right) \\ 
& = & \frac{(1-\alpha_{\rm Py}^2)\sigma_{\rm Py} S}{2e\lambda_{\rm Py}} \mu_{\rm i} =  
\frac{\mu_{\rm i}}{e R_{S}^{\rm Py}}
\end{eqnarray}
This means that in the AP states 
the spin current along the Py pad is injected into the Py particle.  
This discussion also stands for the P configuration as in Fig.\ 4(c).   
Therefore, the spin current induced by non-local spin injection with the negative current injection 
aligns the magnetization of the Py particle along the Py pad.  
The magnitude of the injected spin current can be deduced from the intensity of the 
spin signal by using Eq. (4).  
The relation between the induced spin splitting in the chemical potential 
$\mu_{\rm i}$ at the interface and the obtained spin signal $R_{\rm NLSV}$ in the NLSV measurement 
is given by $\mu_{\rm i} = e i_c R_{\rm NLSV}/\alpha_{\rm Py}$.  
Here, $i_c$ is the exciting charge current for the measurement.  
Therefore, when we inject a pulsed current with an amplitude of $I_{\rm amp}$, 
the injected spin current $I_{S{\rm inj}}$ into the Py particle can be calculated as 
\begin{equation}
I_{S{\rm inj}} = \frac{\mu_{\rm i}}{e R_{S}^{\rm Py}} = \frac{R_{\rm NLSV}I_{\rm amp}}{\alpha_{\rm Py} R_{S}^{\rm Py}}
\end{equation}
As mentioned above, the spin resistance of the Py particle is 0.08 $\Omega$.  
When we use the parameters determined in previous experiments\cite{Kimura3}, 
we obtain the injected spin current as 158 $\mu$A  for $I_{\rm amp} = 14$ mA.   
The value can be compared with that of 
conventional pillar structures consisting 
of Py-based CPP devices.\cite{Py1, Py2}  
The observed spin current for switching the free layer from the AP to P state 
in the vertical structures ranges typically about 200 $\mu$A 
that are comparable to our present experiment.  
As mentioned above, the obtained spin signal is smaller than that 
in the NLSV experiment with the same injector-detector distance.  
We believe that this is caused by the lower quality of the Py particle 
than the Py element in our previous lateral spin-valve experiments.  
In the Py particle, the effect of the surface oxidation is more pronounced than 
the conventional devices because of the small sample dimensions.  
Such effects reduce the spin polarization and the spin diffusion length of the Py particle.  
This causes the reduction of the spin resistance and thus lowers the estimation 
of the spin splitting voltage at the interface.  
The real injected spin current may thus be larger than the above calculation.

Similar analysis for the positive current results in 
the spin current with the same magnitude and the opposite polarity 
injected into the Py particle as in Fig.\ 4(b).  
Therefore, the spin current induced by non-local spin injection 
with the positive current leads the Py particle magnetization into the AP state.  
This transition however is not observed in the present experiment.  
Although the concrete reason has not been clarified yet.  
Conceivable explanations for this discrepancy may be as follows.  
One is a bias dependent spin polarization of the Py pad.  
In general, the spin polarization 
should be independent of the current passing through the interface in the ohmic junction.  
However, it is not true once very thin oxide layer is formed at the interface 
between the Py pad and Cu cross during fabrication process.  
Note that our device is exposed to the air during the process.
Such an oxidized layer may provide an asymmetric barrier especially at high current density.  
Therefore, we have to take into account an asymmetric spin injection into or out of the Py pad.  
When the electron is injected from the Cu wire into the Py pad, 
the spin polarization drastically reduces compared to the zero bias value 
with increasing the applied bias voltage\cite{Monsma}, 
thus diminishing the injected spin current into the Py particle.  
On the contrary, the spin polarization exhibits only small reduction 
when the electron is injected from the Py pad into the Cu wire.  
In this way, our asymmetric reversal of the Py nano-particle can be explained.  
The another possibility is tiny magnetic impurity in the Cu wire near the interface.  
The spin diffusion length of the Cu wire with the magnetic impurity 
is known to depend on the angle between the direction of 
the impurity magnetic moment and that of the injected spin.\cite{Heide}  
When the direction of the injected spin is anti-parallel to 
the moment of the magnetic impurity, the spin diffusion length is shorter 
than that at the parallel alignment because of the reorientation of the magnetic moment 
of the conduction electron spin to the direction of that of the magnetic impurity.  
The magnetic impurity in the Cu wire may be parallel to the magnetization of the Py pad 
because of the exchange interaction.  
In this case, when the electron is injected from the Cu wire to the Py pad (corresponding to 
the positive current), the spin diffusion length is shorter than that of the negative current.  
This asymmetric transport also explains our experimental results.  

In conclusion, we have fabricated the multi-terminal lateral spin injection devices configured for 
the non-local spin-injection-induced magnetization switching of the Py particle.  
We succeeded in switching the magnetization of the Py particle from the AP to the P states by non-local spin injection 
although we could not realize the P to AP switching.  
The conceivable reasons for this discrepancy are discussed.  
The value of the switching current obtained from the experiment was reasonable compared 
with the values estimated from the conventional pillar devices.  
In order to realize both switchings of the Py particle, 
further optimization of the device structure is required.

\begin{figure}

\caption{
Schematic illustrations of (a) the local spin injection 
and (b) non-local spin injection.  
(c) SEM image of the fabricated lateral spin injection device 
and (d) magnified image around the Py particle.  
}

\caption{
(a) Non-local spin valve signal with the probe configuration.  
The solid and dotted lines correspond to the positive and negative field sweep, 
respectively.  
(b) The NLSV signal after the pulsed current injection 
as a function of the current amplitude 
with the the corresponding magnetization configurations.  
}

\caption{
(a) Non-local spin valve signal with the probe configuration.  
The solid and dotted lines correspond to the positive and negative field sweep, 
respectively.  
(b) The NLSV signal after the pulsed current injection 
as a function of the current amplitude 
with the corresponding magnetization configurations.  
}

\caption{
Schematic illustrations of (a) the induced chemical potential with the negative current injection 
in the anti-parallel configuration and 
(b) that with the positive current injection in the parallel configuration.  
(c) Spin-dependent chemical potential inside of the Py particle in parallel (anti-parallel) to 
the Py pad in the negative current.  
The direction of the injected spin current depend on the polarity of the current 
and does not depend on the magnetization configuration between 
the Py pad and the Py particle.  
}

\end{figure}

\newpage

\vspace*{1cm}

\newpage
\vspace*{4cm}
\begin{center}
\includegraphics[scale=0.6]{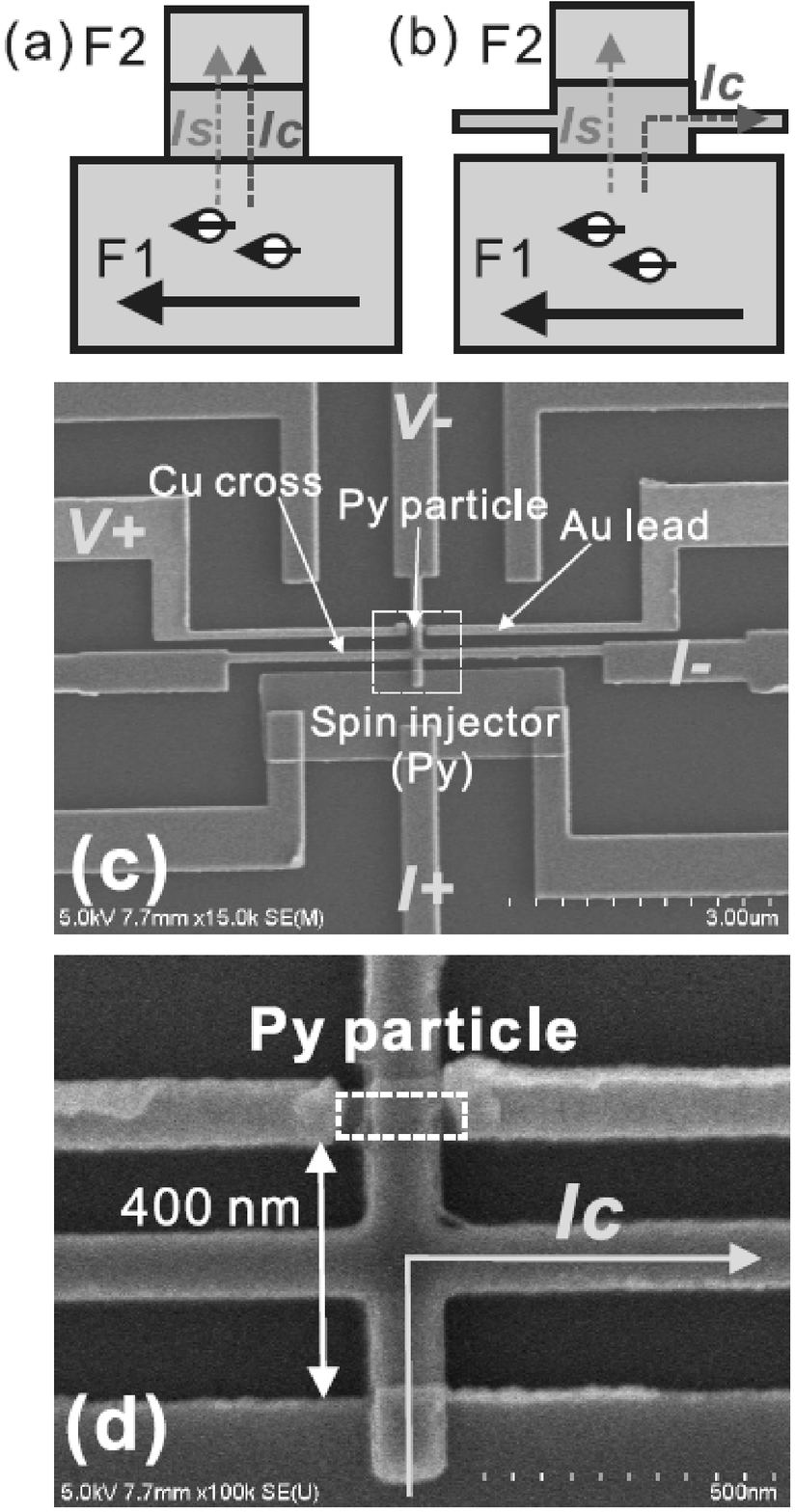}
\end{center}
\vspace*{1cm}
\begin{center}
Fig.\ 1 Kimura et al.
\end{center}

\newpage
\vspace*{2cm}
\begin{center}
\includegraphics[scale=0.45]{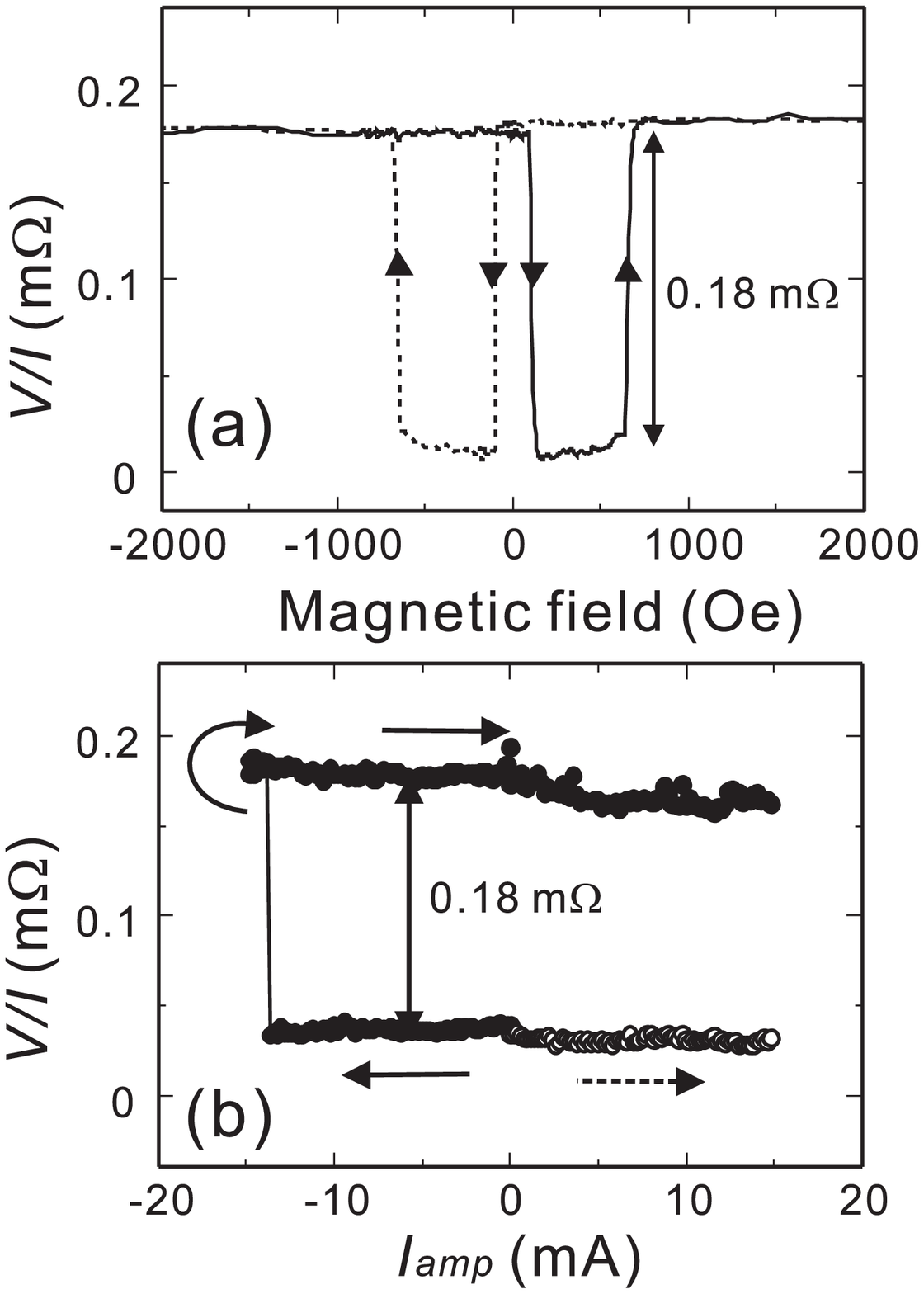}
\end{center}
\vspace*{1cm}
\begin{center}
Fig.\ 2 Kimura et al.
\end{center}

\newpage
\vspace*{4cm}
\begin{center}
\includegraphics[scale=0.45]{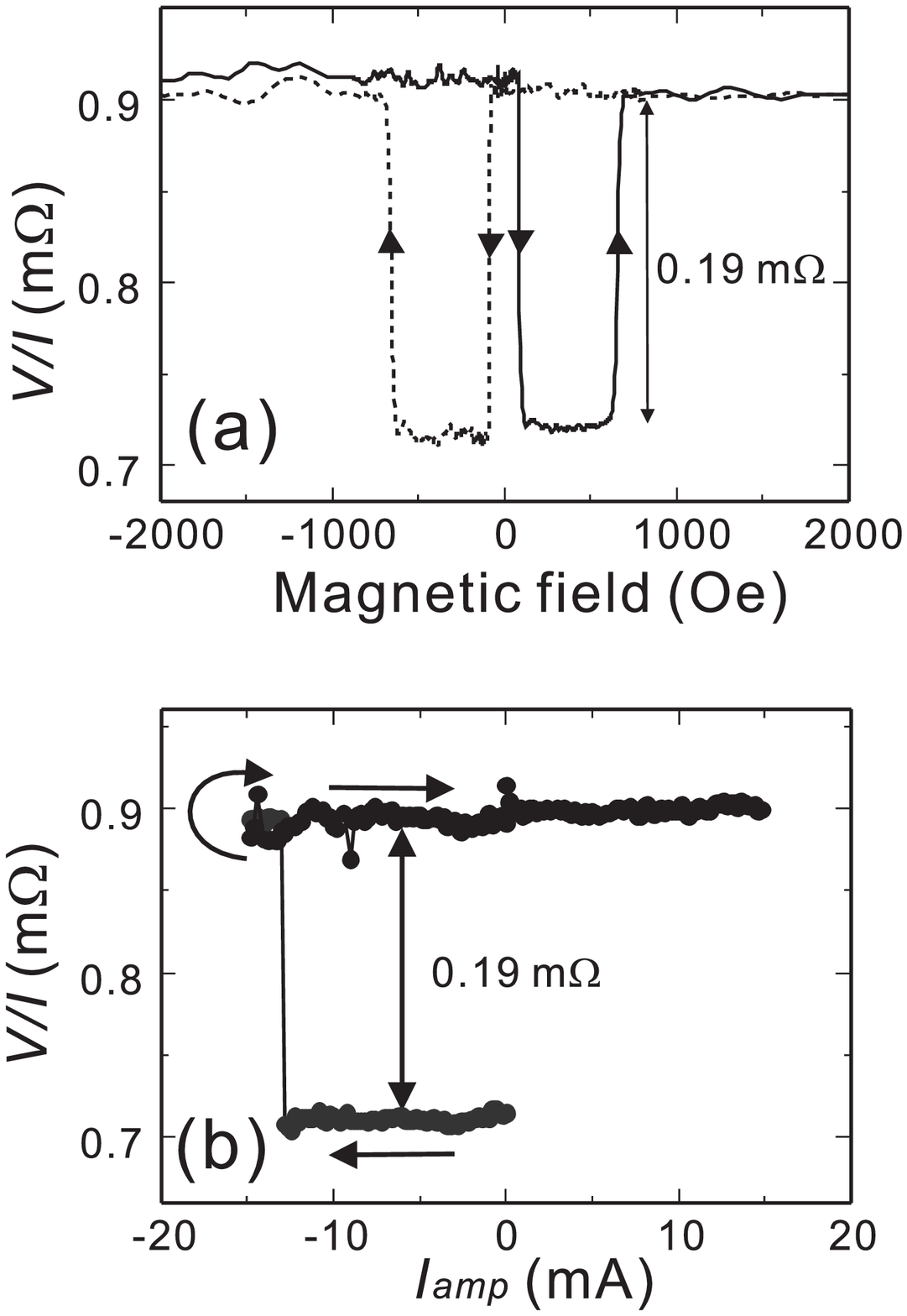}
\end{center}
\vspace*{1cm}
\begin{center}
Fig.\ 3 Kimura et al.
\end{center}

\newpage
\vspace*{4cm}
\begin{center}
\includegraphics[scale=0.7]{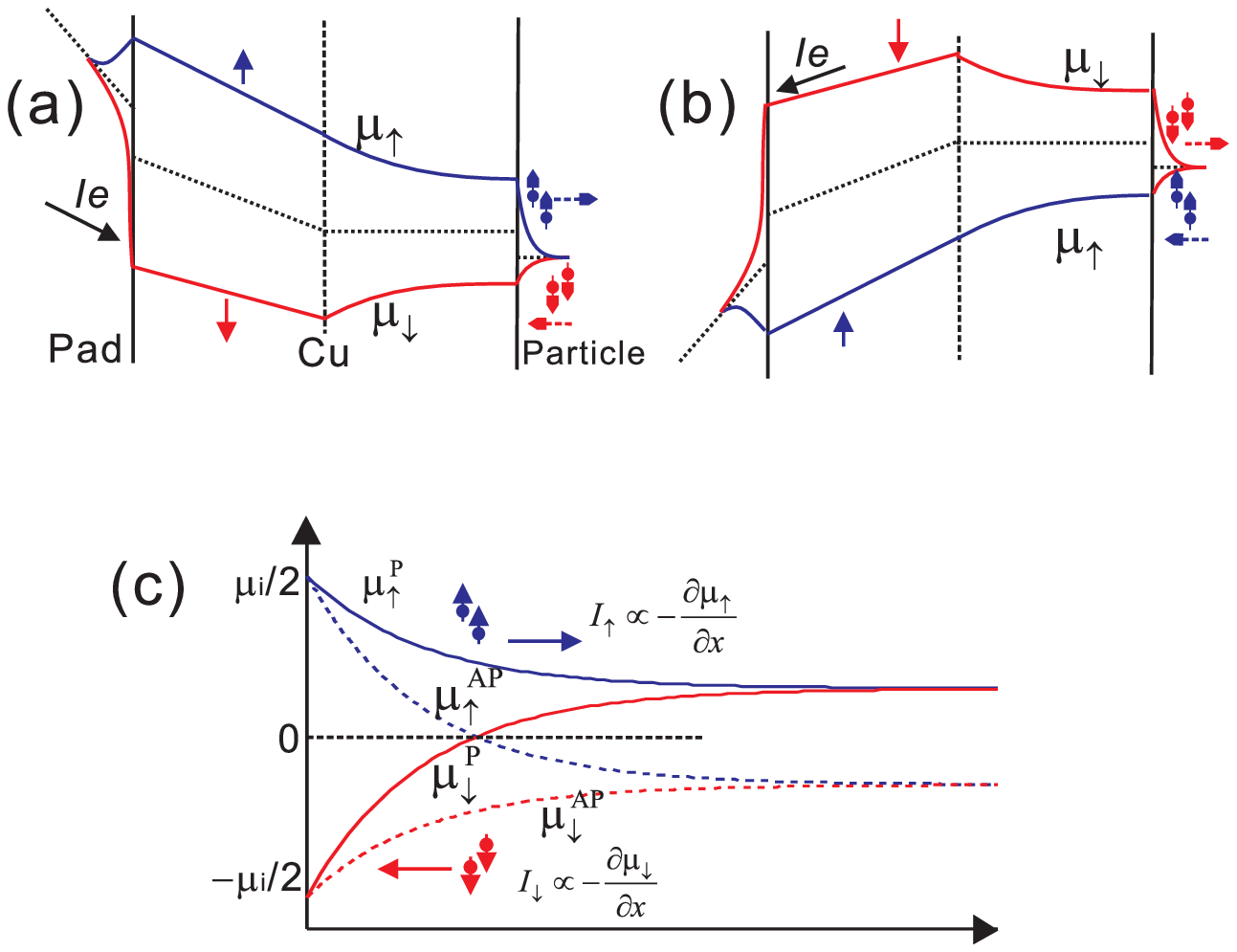}
\end{center}
\vspace*{1cm}
\begin{center}
Fig.\ 4 Kimura et al.
\end{center}

\end{document}